\documentclass[preprint]{aastex}

\def\beq{\begin{equation}}
\def\eeq{\end{equation}}
\def\beqa{\begin{eqnarray}}
\def\eeqa{\end{eqnarray}}

\def\Dwa{$\,$\uppercase\expandafter{\romannumeral5}$\,$}
\def\mic{$\mu$m$\,$}

\def\sles{\lower2pt\hbox{$\buildrel {\scriptstyle <}
   \over {\scriptstyle\sim}$}}

\def\sgreat{\lower2pt\hbox{$\buildrel {\scriptstyle >}
   \over {\scriptstyle\sim}$}}

\begin{document}

\slugcomment{\bf}
\slugcomment{Submitted to Ap.J.}

\title{Calculations of the Far-Wing Line Profiles of Sodium and Potassium in the Atmospheres of Substellar-Mass Objects}

\author{Adam Burrows\altaffilmark{1}, Maxim Volobuyev\altaffilmark{1,2}} 

\altaffiltext{1}{Department of Astronomy and Steward Observatory, 
                 The University of Arizona, Tucson, AZ \ 85721;
                 burrows@zenith.as.arizona.edu, maxim@as.arizona.edu}
\altaffiltext{2}{Department of Chemistry, National Technical University, Kharkiv, Ukraine;
maximnv@vbox.kharkov.ua}

\begin{abstract}

At the low temperatures achieved in cool brown dwarf and hot giant planet
atmospheres, the less refractory neutral alkali metals assume an 
uncharacteristically prominent role in spectrum formation.  In particular,
the wings of the Na-D (5890 \AA) and K I (7700 \AA) resonance lines 
come to define the continuum and dominate the spectrum of T dwarfs
from 0.4 to 1.0 \mic.  Whereas in standard stellar atmospheres the 
strengths and shapes of the wings of atomic spectral lines are rarely 
needed beyond 25 \AA\ of a line center, in brown dwarfs 
the far wings of the Na and K resonance lines out to 1000's of \AA\ detunings are important.
Using standard quantum chemical codes and the Unified Franck-Condon model
for line profiles in the quasi-static limit, we calculate
the interaction potentials and the wing line shapes for the dominant 
Na and K resonance lines in H$_2$- and helium-rich atmospheres.  Our theory 
has natural absorption profile cutoffs, has no free parameters, 
and is readily adapted to spectral synthesis calculations for stars, brown dwarfs, and planets
with effective temperatures below 2000 Kelvin.

\end{abstract}


\section{Introduction}
\label{intro}

Absorption lines of the neutral alkali metals sodium (Na), potassium (K), cesium (Cs), rubidium (Rb), and lithium (Li)
are prevalent in the spectra of L dwarfs, T dwarfs, and irradiated
giant planets (Kirkpatrick et al. 1999,2000; Burgasser et al. 1999,2000,2002; Burrows, Marley, and Sharp 2000 (BMS); 
Burrows et al. 2001,2002; Liebert et al. 2000; Charbonneau et al. 2001; Mart{\'{\i}}n et al. 1999).  
The atmospheres of such substellar mass objects (SMOs)
are cool enough that the neutral alkali atoms predominate.
Furthermore, since aluminum, magnesium, iron, silicon, and 
calcium are sequestered in grains that have settled out,
a cool atmosphere in the gravitational field of such a planet or star is 
depleted of these more refractory metals (Burrows and Sharp 1999; Lodders 1999; BMS).
As a consequence, the less-refractory neutral alkali
metals assume an importance in the optical and near-infrared
spectra of SMOs and in their atmospheres from $\sim$800 K to 
$\sim$2000 K that is unique among astronomical objects.   

In standard stellar atmospheres, atomic lines are superposed on
a background continuum and the concepts of equivalent width and 
curve of growth make conceptual and practical sense.  An individual
line is but a perturbation on the local spectrum.  However, due to
the rainout of metals in cool atmospheres and the consequent 
paucity of continuum and alternate opacity sources 
between 0.4 \mic and 1.0 \mic, the wings of the strong resonance doublets
centered at $\sim$7700 \AA\ (K I) and 5890 \AA (Na-D) 
assume the role of the continuum throughout most of this
spectral range (BMS; Tsuji, Ohnaka, and Aoki 1999).  
In particular, the red wing of  
the $4s^2S_{1/2}-4p^2P_{3/2}$ transitions of K I provides 
the pseudo-continuum in cool molecular atmospheres 
all the way from 0.77 \mic to $\sim$1.0 \mic and the Na-D doublet, centered
as it is in the middle of the visible, determines the true color of brown dwarfs
(magenta/purple; Burrows et al. 2001).

Hence, whereas in traditional stellar atmospheres the
Lorentzian core and Gaussian wings of a line are not generally of relevance beyond    
$\sim$20 \AA\ detunings ($\Delta\lambda$ from the line core), in cool substellar
atmospheres the relevant reach of the Na I and K I resonance lines can be
thousands of \AA.  Given this, to achieve accurate spectral fits   
for brown dwarf, L dwarf, T dwarf, and hot giant planet atmospheres,
the shapes of the far wings of these alkali lines as a function of
pressure and temperature must be ascertained.   

In this paper, we perform ab initio calculations of the energy shifts
of the ground and excited states of sodium and potassium 
immersed in H$_2$- and helium-rich 
atmospheres and obtain the resulting opacity profiles of the red and blue wings
of the Na-D and K I resonance lines.
In \S\ref{quasi}, we review the general theory of line profiles in the
quasi-static limit using the Unified Franck-Condon (UFC) formalism (Szudy and Baylis 1975,1996).  
In \S\ref{technique}, we describe our use of the quantum chemical
code GAMESS (Schmidt et al. 1993) to obtain the interaction potentials as a function
of H$_2$ and helium perturber distance and H$_2$ orientation. The derived potential curves are presented
and described in \S\ref{potential}, where we also compare to related work in the literature.  The resultant cross
sections as a function of wavelength are given in \S\ref{cross}.  The latter section contains our major results. 
We compare our new theory with the approximate theories found in Burrows, Marley, and Sharp (2000)
and Burrows et al. (2002) and find both similarities and important differences.
In \S\ref{conclusion}, we summarize our major conclusions.

\section{Quasi-Static Theory of Absorption Spectra in the Far Wings of Atomic Spectral Lines}
\label{quasi}

The energy levels of an atom of sodium or potassium immersed in a sea of molecular hydrogen will
be perturbed by the potential field of the diatomic hydrogen.  At a distance of $R_i$, the molecular
hydrogen will shift both the ground and the excited states of the alkali metal atom.
The difference of these level shifts is the amount by which the corresponding absorption
line will be shifted (Griem 1964; Breene 1957,1981):
\begin{equation}
h\nu = h\nu_0 + V_m(R_i) - V_n(R_i) = h\nu_0 + \Delta V_{mn} (= \Delta V)\, ,
\label{shift}
\end{equation}
where $\nu$ is the photon frequency, $\nu_0$ is the unperturbed photon frequency,
$m$ and $n$ indicate the excited and ground states, and for an H$_2$ perturber
the potential shifts $V$ are a function of the orientation (angle) of the molecule and the distance.
Note that $V_m(R_i)$ and $V_n(R_i)$ are the potential shifts with respect 
to the asymptotic shifts at infinity.
$\Delta V$ is a central quantity in the quasi-static theory of line profiles.

The quasi-static theory of absorption strength assumes that the perturbers are arrayed
about the transitioning atom in a static, thermal distribution with a density $N_p$
given by its partial pressure and temperature ($T$) (Holtzmark 1925; Holstein 1950).  
A full quantum mechanical treatment of line shapes has been developed, but it has been applied
to only a limited number of systems and is significantly more involved (Szudy and Baylis 1996).
Fortunately, at least for the Na-He system, Pontius and Sando (1983) have shown that the
semi-classical/quasi-static and the quantal methods yield very similar results.   

In the quasi-static theory one imagines a shell of radius $R_i$ and 
thickness $dR_i$ centered at the atom.  The strength of the absorption line in the wings at the
frequency given by eq. (\ref{shift}) is then proportional to the number of perturbers in that shell:
\begin{equation}
4\pi R_i^2 N_p dR_i = \frac{4\pi R_i^2}{d\Delta V/dR_i} N_p d\Delta V\, .
\label{simple}
\end{equation}
The quantity $d\Delta V/dR_i$ has been extracted to show $d\Delta V$, which is proportional
to the differential frequency ($d\nu$; eq. (\ref{shift})).  Eq. (\ref{simple}) indicates  
that the absorption cross section at frequency $\nu$ is proportional to:
\begin{equation}
\frac{4\pi R_i^2}{d\Delta V/dR_i} N_p e^{-\frac{V_0(R_i)}{k_BT}}\, ,
\label{simple2}
\end{equation}
where $k_B$ is Boltzmann's constant, $V_0(R_i)$ is the ground state energy shift by a 
perturber at distance $R_i$, and the density of perturbers has been corrected by the 
corresponding thermal Boltzmann factor.   Inverting eq. (\ref{shift}) 
to obtain $R_i(\nu)$ at a given absorption frequency
yields an expression for the absorption cross section in the
line wings at that frequency.  Each solution for $R_i$ at a given $\nu$ is called a Condon
point, where the subscript $i$ is now revealed to be the Condon point index.  
In principle, since $\Delta V$ can be multiple-valued, at a given frequency there can be multiple
Condon points.  Expressions like eq. (\ref{simple2}) must be summed
for all such points to obtain the total cross section.  

Expression (\ref{simple2}) is not only missing the dipole transition moment (oscillator strength $f_{mn}$),
but is not normalized by the dipole (Thomas-Reiche-Kuhn) sum rule to yield the correct cross section $\sigma(\nu)$. 
Furthermore, it must be corrected at small detunings (large $R_i$).  The correct expression
for the absorption cross section in the wings of a spectral line in the quasi-static approximation is given 
by the Unified Franck-Condon formula (Szudy and Baylis 1975,1996):
\begin{equation}
\sigma(\nu) = \sum_i \sigma_i(\nu) + \sum_{i,j=i+1} \sigma_{ij}(\nu)\, ,
\label{total}
\end{equation}
where
\begin{equation}
\sigma_i(\nu) = A\frac{R_i^2}{|\Delta V^{\prime}(R_i)|}\bigl(\pi z_i\bigr)^{1/2}\Biggl(L(z_i) + \frac{M(z_i)}{z_i}\Biggr)e^{-\frac{V_0(R_i)}{k_BT}}
\label{sigmaufc}
\end{equation}
and
\begin{equation}
\sigma_{ij}(\nu) = 2A\frac{R_i^2}{|\Delta V^{\prime}(R_i)\Delta V^{\prime}(R_j)|^{1/2}}\bigl(\pi Z_{ij}\bigr)^{1/2}\Biggl(L(Z_{ij}) - \frac{M(Z_{ij})}{Z_{ij}}\Biggr)e^{-\frac{V_0(R_i)}{k_BT}}\, .
\label{rainbow}
\end{equation}
$A$ contains the transition oscillator strength and perturber number density and is given by:
\begin{equation}
A = 12\pi N_p\frac{\pi e^2}{m_ec^2}f_{mn} \, .
\label{AAA}
\end{equation}
The arguments $z_i$ and $Z_{ij}$ are given by the expressions:
\begin{equation}
z_i = \frac{1}{2}\Biggl(\frac{\mu}{k_BT}\Biggr)^{1/3}\Biggl(\frac{\Delta V^{\prime}(R_i)}{\hbar}\Biggr)^2\Biggl\vert \frac{\Delta V^{\prime\prime}(R_i)}{\hbar}\Biggr \vert^{-4/3}
\label{zdef}
\end{equation}
and 
\begin{equation}
Z_{ij} = \frac{1}{2}\Biggl(\frac{\mu}{k_BT}\Biggr)^{1/3}\Biggl(\frac{\Delta V^{\prime}(R_i)\Delta V^{\prime}(R_j)}{\hbar^2}
\Biggr)\Biggl\vert \frac{\Delta V^{\prime\prime}(R_i)\Delta V^{\prime\prime}(R_j)}{\hbar^2}\Biggr \vert^{-2/3}\, ,
\label{bigzdef}
\end{equation}
where
\begin{equation}
\Delta V^{\prime}(R_i) = \frac{\partial \Delta V}{\partial R_i}\, 
\end{equation}
and $\mu$ is the reduced mass of the interacting pair.
$L(z_i)$ and $M(z_i)$ are functions defined through the integrals:
\begin{equation}
L(z_i) = \int^{\infty}_0 d\zeta\zeta^{-2}Ai^2(-z_i\zeta)e^{-\zeta^{-3}}
\end{equation}
and 
\begin{equation}
M(z_i) = \int^{\infty}_0 d\zeta\zeta^{-3}Ai^{\prime 2}(-z_i\zeta)e^{-\zeta^{-3}}\, ,
\end{equation}
where $Ai(x)$ is the Airy function and $Ai^{\prime}(x)$ is its derivative.  
The asymptotic behavior of $L(z_i)$ and $M(z_i)$ for positive $z_i$ is:
\begin{equation}
L(z_i) = \bigl(36\pi z_i\bigr)^{-1/2} 
\end{equation}
and
\begin{equation}
M(z_i) = \Bigl(\frac{z_i}{36\pi}\Bigr)^{1/2}\,  .
\end{equation}

Eq. (\ref{sigmaufc}) is the expression that corresponds to eq. (\ref{simple2}), the 
sum is over the Condon points ($i$ and $j$), and $\sigma_{ij}(\nu)$ is the interference term between the Condon points.
The interference term is important only when $R_i \sim R_j$, at which point $\Delta V^{\prime}(R_i)$ is formally
zero.  This singularity (seen in eqs. (\ref{simple2}) and (\ref{sigmaufc})) yields what is referred to as a ``rainbow" 
or ``satellite" feature (Beuc and Horvatic 1992) in the wings of an absorption spectrum.  We expect such
features in the Na I (5890 \AA) and K I (7700 \AA) absorption cross sections (\S\ref{potential} and \S\ref{cross}).     
Note that at large $z_i$, the expression $\bigl(\pi z_i\bigr)^{1/2}\bigl(L(z_i) + \frac{M(z_i)}{z_i}\bigr)$ in
eq. (\ref{sigmaufc}) becomes a constant and eq. (\ref{sigmaufc}) reduces to eq. (\ref{simple2}), modulo constants
and the oscillator strength.  Moreover, at large $Z_{ij}$, the expression $\bigl(\pi Z_{ij}\bigr)^{1/2}\bigl(L(Z_{ij}) - \frac{M(Z_{ij})}{Z_{ij}}\bigr)$
in eq. (\ref{rainbow}) goes to zero. 

The exponential terms in eqs. (\ref{sigmaufc}) and (\ref{rainbow}) provide natural 
cutoffs at positive (red) and negative (blue) detunings and depend
only upon the ground-state potential shifts and temperature.  Such cutoffs have been shown to be necessary
for reasonable spectral fits to T dwarf spectra longward of $\sim$1.0 \mic (Burrows et al. 2002) and are
naturally provided by the quasi-static UFC theory.

The line cores (not the subject of this paper) are determined by distant
encounters and are handled by assuming a van der Waals interaction
potential with an adiabatic impact theory (Weisskopf 1933; Ch'en and Takeo 1957;
Dimitrijevi\'{c} and Peach 1990), as in BMS and Nefedov, Sinel'shchikov,
and Usachev (1999).  The result is a pressure-broadened core of Lorentzian shape that
applies out to only tens of wavenumber (cm$^{-1}$) detunings.  The major contribution to the
alkali line profile shape is determined by the statistical UFC theory, which 
applies beyond the core to thousands of wavenumber (and thousands of Angstrom) detunings.
The procedure for obtaining the wing absorption profiles is first to calculated the potential curves
for the relevant levels of Na and K (\S\ref{technique}  and \S\ref{potential}), then to calculate the Condon points, and finally to
employ eq. (\ref{total}).   We do this for Na+H$_2$, Na+He, K+H$_2$, and K+He interaction pairs,
and for the states that contribute to both the red and the blue wings of the 5890 \AA\ and 7700 \AA\ resonance features. 
For perturbations by H$_2$, the potential interactions are functions of angle (orientation), 
introducing yet another degree of complexity.

\section{Technique for Calculating the Interaction Potentials and Line Shifts}
\label{technique}

The Na-D doublet of neutral sodium involves the $3s^2S_{1/2}-3p^2P_{3/2,1/2}$ transitions, 
where the D$_2$ transition (5890 \AA) is from the $1/2$ state to the $3/2$ state with a multiplicity
of 4 ($=2j + 1$) and the D$_1$ transition (5896 \AA) is from $1/2$ to $1/2$ with a multiplicity
of 2.  Hence, the ratio of the D$_2$ and D$_1$ line strengths is 2.  The analogous transitions
for potassium are $4s^2S_{1/2}-4p^2P_{3/2,1/2}$ at 7665 \AA\ and 7699 \AA, respectively,
with a similar strength ratio. Since the upper states of the D$_2$ transitions are $j=3/2$ states, 
they include $p_x$ and $p_y$ states that in molecular orbital
theory can form bonding states with H$_2$ (oriented perpendicularly).  The $p_z$ state associated 
with the $^2P_{1/2}$ state is anti-bonding to H$_2$.
Figure \ref{fig:1} depicts the orientations of the interacting H$_2$ and Na pair for C$_{2v}$
symmetry (H$\rightarrow$H line perpendicular to the Na-H$_2$ line).  The potassium $4p$ states can be depicted 
in a perfectly analogous way. The $^2A_1$ states depicted in Fig. \ref{fig:1}
are anti-bonding and the interaction potential is purely repulsive at all distances $R_i$.
The $^2B_{1,2}$ states are attractive at large distances and repulsive at small distances
\footnote{Note that in order to connect to both communities, we use both the spectroscopist's ($A$ and $B$) and the 
physicist's ($P_{1/2,3/2}$) nomenclature for the states.}.  The ground states 
of both the Na+H$_2$ and the K+H$_2$ systems are repulsive for all distances
$R_i$.  The upshot is that the energy difference between the $^2$B state of the D$_2$ transition
and the ground state decreases with decreasing $R_i$ until $R_i$ is quite small.  Hence, the D$_2$ line,
though it is blueward of the D$_1$ line for atoms in isolation, provides the red wing
of the alkali line.  Correspondingly, the D$_1$ line provides the blue wing.  

A schematic of the excited-state minus ground-state energy shifts of the Na-D and K I resonance lines
is given in Fig. \ref{fig:2}.  Figure \ref{fig:2} depicts the mapping between energy shift, line core, red wing, and blue wing.  
Note that $\Delta V$ on the blue wing has a maximum (extremum), which results in a rainbow satellite feature.
The absence of a corresponding extremum on the red wing means that there is no corresponding satellite feature 
redward of the line core.  

To calculate the interaction potentials, we employ the multi-configurational 
self-consistent-field (MCSCF) variant of the quantum chemical code GAMESS (Schmidt et al. 1993).
The Hartree-Fock (HF) method is a self-consistent field approach that employs 
iteration to find a consistent solution to the multi-electron, multi-nucleus wavefunction (itself a Slater determinant),
but assumes that the potential is a smoothed average and does not explicitly include electron-electron interactions.  
MCSCF is a post-HF method that includes electron-electron interactions 
and exchange forces, the so-called configuration interaction (CI), but assumes that 
the CI term is small.  It allows variation of not only the mixing coefficients of the various HF configurations,
but also of the coefficients of the basis functions in the constituent molecular orbitals.  In this way, 
the HF orbitals from which the total wavefunction is constructed are optimized simultaneously.  
The MCSCF method requires care in the selection of the basis set, but generally
achieves a good solution for problems having low-lying excited states, as 
in the problem at hand.  We use the Complete Active Space Self-Consistent Field (CASSCF)
version of MCSCF and the 6-31G$^{**}$ split-valence basis set that 
incorporates Gaussian Type Orbitals (GTO) (Frisch et al. 1998).
In practice, we first obtain a solution with the HF method, and then use this solution
as a first guess in the full MCSCF/CASSCF calculation.  This two-step approach
has often proven necessary to achieve converged solutions.

We use the above methods and the GAMESS code to solve the many-electron Schr\"odinger equation
for the wavefunctions and the energy levels of the various electronic 
states as a function of distance $R_i$ for the composite 
system of alkali metal atom and either H$_2$ or helium.  
For H$_2$, this is done for a variety of orientations
(from C$_{2v}$ [$\theta = 90^{\circ}$] to C$_{\infty}$ 
[$\theta = 0^{\circ}$] (parallel) symmetries) and we
keep the H--H bond length constant (Botschwina et al. 
1981).  We have included electron correlation
effects successfully for He 
perturbations and for H$_2$ perturbations germane to the red wings of
both the Na and K lines.  However, when calculating the blue wings for 
orientation angles other than 0$^{\circ}$ or 90$^{\circ}$  
we had difficulty obtaining converged solutions with 
the electron correlation effects (CI) included.  Hence, for the blue wings 
of the Na/K+H$_2$ interaction pairs we have settled for the pure Hartree-Fock solutions.
This approximation is good for large $R_i$, but begins to break down at small $R_i$ and large detunings. 

For all the relevant $^2A$ and $^2B$ states we calculate the interaction potentials 
as a function of both distance and (for H$_2$) angle of orientation.
The calculations are done at hundreds of distances and (for H$_2$) for five angles 
($\theta = 0^{\circ}, 20^{\circ}, 45^{\circ}, 70^{\circ},$ and $90^{\circ}$).
Then, we calculate the energy shifts ($\Delta V$) of the 
resonance transitions of Na and K.  Though we reproduce the wavelengths of the isolated D$_2$ and D$_1$
lines of Na to within $\sim$14\% (0.3 eV) and for K to within $\sim$21\% (0.34 eV), 
we shift our calculated energy shifts by a constant amount that ensures that 
the line shift at very large $R_i$ is zero.  This puts the line core exactly at the measured
wavelength by construction.  Furthermore, we do not calculate the dipole transition matrix elements 
as a function of $R_i$, but use the published oscillator strengths of the unperturbed Na and K 
resonance doublets in the calculation of the line profiles (eq. \ref{AAA}).  These incorporate
the 2:1 strength ratio due to the spin degeneracy of the final states in the unperturbed 
D$_2$ and D$_1$ transitions.

\section{Potential Curves for the Na+H$_2$, K+H$_2$, and Na/K+He Systems}
\label{potential}

Figure \ref{fig:3} depicts the energy levels as a function of separation 
of the various states (relative to the ground state at infinite separation),
for both Na and K perturbed by both helium and H$_2$.  For H$_2$, the run of energy shift with $R_i$ for the four states 
is given for various orientation angles.  As Fig. \ref{fig:3} shows, the ground state interaction
is always repulsive and is most repulsive for H$_2$ for colinear orientations (C$_{\infty}$ symmetry).
This is the $V_0(R_i)$ term to be used in eqs. (\ref{sigmaufc}) and (\ref{rainbow}) in calculating the cutoff exponential.
The $^2B$ states associated with the $p_x$ and $p_y$ atomic orbitals of sodium or potassium 
show a slight attraction for large angles (close to perpendicular) and large distances, but then become universally repulsive
as the hard core is approached (Botschwina et al. 1981).  Figure \ref{fig:4} portrays the resulting transition energies 
for the D lines of both Na and K as a function of separation and angle.  As is clear from the figure,
the energy differences for the transitions for which the excited states are $^2B$ states decrease
with decreasing separation and, hence, are associated with the red wings of the spectral lines.
Similarly, since the ground-state to $^2A_1$-state transitions have the opposite behavior, 
they are associated with the blue wings.   Furthermore, the peaks seen at the bottom of Fig. \ref{fig:4}
at separations of $R_i \sim 2.4-2.5$\AA\ and $R_i \sim 3.1-3.2$\AA, respectively,
for Na and K perturbed by H$_2$ imply that their blue wings have satellites 
at wavelengths of $\sim$0.49-0.5 \mic and $\sim$0.66--0.68 \mic.
The K satellite might be near the Li feature at 6708 \AA.  Fig. \ref{fig:2} schematically shows
how the panels of Fig. \ref{fig:4} merge to make the corresponding absorption profiles. 

It is useful to compare these potential calculations with previous related work
using different methods.  Rossi and Pascale (1985, RP) have done $l$-dependent 
pseudopotential calculations for the Na+H$_2$ and K+H$_2$ systems
and Botschwina et al. (1981) have done RHF-SCF and PNO-CEPA calculations for the Na+H$_2$ system.
The PNO-CEPA potential curves of Botschwina et al. are similar to those of RP. 
Neither did the calculations for the arbitrary H$_2$ orientations 
and small separations necessary to perform the line 
profile calculations that motivated this paper.  Rather, 
they concentrated solely on separations greater than 
2.0 \AA\ and on the C$_{\infty}$ and C$_{2v}$ 
symmetries.  Nevertheless, their potential curves for these symmetries are 
similar to those found in Fig. \ref{fig:3}.  For C$_{2v}$ 
symmetry and the Na+H$_2$ system, RP derive that the potential minimum
for the $^2B_1$ state is at 2.4 \AA\ and has a depth of 0.12 eV; we obtain 2.45 \AA\ and 0.1 eV.  For the
$^2B_2$ state, RP derive corresponding numbers of 2.12 \AA\ and 0.35 eV; we obtain 2.27 \AA\
and 0.2 eV.  For the K+H$_2$ system, RP derive for the $^2B_2$ state a minimum at 3.0 \AA\
with a depth of 0.49 eV; we obtain 3.0 \AA\ and 0.5 eV.  RP estimate that the change in the potential of the 
$^2A_1$ excited state of the Na+H$_2$ system with C$_{2v}$ symmetry in going from 4.0 Bohr radii (2.12 \AA) to
8.0 Bohr radii (4.23 \AA) is 0.55 eV; we obtain 0.7 eV.  Furthermore, Kleinekath\"ofer et al. (1996),
who use a surface integral method to derive the ground-state potential curve of the Na+He system, derive
V($R_i$ = 3.2 \AA) = 0.10 eV; we obtain 0.08 eV at the same distance.  All in all, our results compare 
favorably with those in previous studies, with the Na+H$_2$/$^2B_1$ and K+H$_2$/$^2B_2$ potential
curves corresponding most closely.  However, the correspondences are not perfect and we sometimes see
discrepancies of as much as $\sim$0.1 eV at a given separation.  We suspect that these discrepancies are traceable 
either to our choice of basis sets or to the MCSCF assumption that the CI term is small.

Since the second derivative of the potential ($V^{\prime\prime}$)
is needed when using eqs. (\ref{sigmaufc}) and (\ref{rainbow}) to obtain the satellite profile shapes,
small errors in the potential calculations when $V^{\prime} = 0$ can be amplified at the rainbow positions.
Ch'en and Wilson (1961) measured the ``violet band separation" for the satellite features on the blue wings 
of the Na+He and the K+He systems to be 1161$\pm$15 cm$^{-1}$ and 838$\pm$5 cm$^{-1}$, respectively,
which translate into satellite positions at 0.55 \mic and 0.72 \mic.  These contrast with
the satellite positions we derive of 0.515 \mic and 0.69 \mic, respectively, for perturbation by helium.
Not unexpectedly, the centers of the satellite features are the most poorly determined quantities in our theory. 
We now turn to the calculation of the opacity profiles in the wings of the D$_2$ and D$_1$ lines of Na and K
due to perturbations by H$_2$ and helium using the potential curves represented in Figs. \ref{fig:3} and \ref{fig:4}.

\section{The Absorption Cross Sections on the Red and Blue Wings of the Na (5890\AA) and K (7700\AA) Resonance Lines}
\label{cross}

Using the UFC/quasi-static eqs. (\ref{sigmaufc}) and (\ref{rainbow}), we obtain
the absorption spectra as a function of photon wavelength for the D$_2$ and D$_1$ 
lines of Na and K centered near 5890 \AA\ (0.589 \mic) and 7700 \AA (0.77 \mic),
respectively.  These are depicted in Fig. \ref{fig:5} at a temperature of 1000 K
and a total pressure of one atmosphere.  The contributions at 
various orientation angles of H$_2$, the angle-integrated spectra (solid),
and the results for the Na+He and K+He systems are shown.  The Lorentzian cores 
are not included on this plot.  Since the ground-state potential shift increases quickly 
at small separations (Fig. \ref{fig:3}), and the associated exponential terms in eqs. ({\ref{sigmaufc})
and (\ref{rainbow}) introduce a large cutoff, Condon points at such small separations make only a 
marginal contribution to the overall opacities.  As a consequence, though
there can be more than two Condon points (Fig. \ref{fig:4}), we include at most  
the outer two in our theory.  For the same reason, we ignore possible satellites
on the very far red wings.

Figure \ref{fig:5} encapsulates the central results of our study.  The cutoff on the red
wing of the potassium feature due to the exponential term (eq. \ref{sigmaufc}) containing the ground-state
interaction potential of the K+H$_2$ system is seen to be situated between 0.95 \mic 
and 1.0 \mic, close to the 0.98 \mic used in Burrows et al. (2002).  The cutoff of the
red wing of the Na feature is near 0.8 \mic, but is a bit more gradual.  The corresponding 
cutoffs for the alkali-He systems are more abrupt, but of less importance due to the lower abundance of helium.
It should be noted that for the temperatures in cool atmospheres the precise form and position
of this cutoff is a sensitive function of the ground-state potential.  
Errors in the potential of \sles{0.1 eV} that may arise at small $R_i$ for the alkali-H$_2$
systems translate directly into shifts in the position of the turnover and in the 
magnitude of the cross section at the turnover.  Such an effect  
is demonstrated by the difference between the cross sections 
with and without electron-correlation/configuration-interaction terms (depicted below in Fig. \ref{fig:7}).   
Figure \ref{fig:5} also makes clear that the red and blue wings are asymmetrical, as one would expect
from the discussion in \S\ref{potential} and Fig. \ref{fig:4}.

The satellite features identified on the blue wings of both the Na and K features by the asterisks on Fig. \ref{fig:5}
have been alluded to before, but we note once again that our calculations are too
imprecise to derive the detailed shape of the actual rainbow structures (which for the alkali-H$_2$ systems 
is a composite of the corresponding curves for the various angles).  Nevertheless, the 
existence and approximate wavelengths of the satellites remain clear.

Figure \ref{fig:6} portrays the temperature and pressure dependences of the line profiles. 
Three curves (at 800, 1000, and 1500 K) show the temperature dependence at one
atmosphere pressure and one curve at 1000 K and 10 atmospheres shows, by comparison with the other 1000 K curve, the  
pressure dependence.  Due to the $N_p$ dependence in eq. (\ref{AAA}), higher pressures lead to higher wing strengths.
The Thomas-Reiche-Kuhn sum rule is still satisfied by a corresponding diminution in the strength
in the Lorentzian core.  Due to the exponential term, increasing the temperature raises the cross section in the
far wings.  However, due to the constraint of constant pressure, increasing the temperature slightly lowers 
the cross sections in the near wings. 

It should be emphasized that, unlike the theory of BMS,
and the line profile formalism used in Burrows et al. (2002), 
there are no free parameters in our theory.  The cutoffs
and shapes arise naturally because the physics of the
line-broadening process has been treated.  Limitations on the
accuracy of the current work are in the accuracy of the quantum
chemistry calculations of the potential curves and in the use
of the semiclassical formalism to treat the broadening.
The earlier theory of BMS was constructed to fill the theoretical and experimental
void concerning the far wing profiles of the alkali metal lines perturbed by molecular 
hydrogen.  Burrows et al. (2002) followed with a simpler theory that used 
a Gaussian-truncated Lorentzian profile.  Previously in astrophysics  
the character of the far wings of the K I and Na D
resonance features perturbed by H$_2$ had not been needed.

Given our new calculations, it is
useful to compare them with the stopgap theories of BMS and Burrows et al. (2002).
Figure \ref{fig:7} is a comparison 
of the BMS (red) and Burrows et al. (2002) (green) potassium cross sections with 
the corresponding cross sections using the theory of this paper (blue).   For the BMS prescription,
we arbitrarily set the associated $q$ parameter equal to 0.5.  A temperature
of 1000 K and a pressure of one atmosphere were assumed.  The BMS and the Burrows et al. (2002) curves include the
Lorentzian core, while the newly-calculated quasi-static curve does not.  Also shown are the new cross sections calculated 
without the configuration interaction (CI) (blue, dashed). 
At the inner red wing, the theory of Burrows et al. (2002),
in which the pressure-broadened Lorentzian was
truncated with a Gaussian near 0.98 \mic,
is within a factor of three of the more detailed
calculation, but clearly deviates from it.  However, on 
the red wing the simple theory of BMS 
is actually surprisingly close to the new result.    
On the far red wing, the cross sections of 
the new theory plummet more precipitously than
either BMS or Burrows et al. (2002).  This emphasizes 
even more strongly the existence of the steep cutoff
inferred by Burrows et al. (2002) to be necessary in order 
to reproduce the $Z$ ($\sim 1.0$ \mic) and $J$ ($\sim 1.2$ \mic)
band fluxes observed for the T dwarfs (Burgasser et al. 
2002).  Note, however, that the new cross sections
on the far blue wing are as much as a factor of four  
higher than the corresponding cross sections  
of the old theories.  The same result obtains for 
the blue wing of the Na-D line.  In sum, though the 
BMS theory was good in the near red wings, in places the contrast
between both the BMS and the Burrows et al. (2002) formulations 
and our parameterless UFC calculations is large.
As a result, we recommend that in the future the
new theory be used for detailed T and L dwarf spectral calculations.

\section{Conclusions}
\label{conclusion}

Using the quantum chemical code GAMESS, we have calculated 
as a function of distance the interaction potentials of the excited
and ground states of sodium and potassium perturbed by H$_2$ and helium.
For H$_2$, we also calculated the dependence of the interaction potentials on orientation angle.
Using these potentials and the Unified Franck-Condon formalism, we derived the 
absorption line profiles in the wings of the Na-D doublet and the K I resonance doublet
at detunings of thousands of Angstroms.
These lines and their wings are of central importance in the atmospheres and spectra 
of brown dwarfs, T dwarfs, L dwarfs, and hot Jupiters.  Our theory has no free parameters
and naturally accounts for the cutoffs in the far red wings inferred to
be present to explain the high $J$ and $Z$ band fluxes observed in T dwarfs (Burrows et al. 2002).
However, the new cross sections differ from those used previously by having sharper
cutoffs, stronger blue wings, and satellite features.
Furthermore, the red/blue asymmetry is pronounced.  

From our calculations using the full configuration interaction at
the subset of useful H$_2$ orientation angles for which we were able 
to obtain convergence for the $^2A_1$ states (of relevance for the calculation of the blue wings)
and from a comparison of the Hartree-Fock and CI results for perturbations by helium,
we have determined that the HF and CI numbers on the blue wings differ little, though
the associated potential differences on the red wings can differ by $\sim$0.1 eV.  This gives us some confidence that,
despite the fact that we use the HF approach on the blue wings for perturbations by H$_2$,
our blue wing results are satisfactory.  Fortunately, for all the red wing results, we were able to obtain
full convergence using the MCSCF/CI formalism at all H$_2$ angles.  

The need for line wing strengths at such large
detunings has few precedents in astrophysics.
The next step is to incorporate the new alkali line opacities into a spectral synthesis
code and to identify the differences between the fits using the old and the new algorithms.
The line cores will be unchanged, but the blue wings and the spectra from 0.8 \mic to 0.9 \mic
on the red wings of the K I line will be slightly altered, as else being equal.  The major virtue
of our new algorithm is that it is parameterless.  As a result, we reduce by one (or even a few)
the number of free parameters needed in fits to the increasing number of optical and near-IR spectra
being measured for T dwarfs (in particular), thereby making the extraction of 
physical quantities such as effective temperature 
and gravity more robust and reliable.

\acknowledgments

We are happy to thank Ludwik Adamowicz, Jonathan Lunine, Christopher Sharp, David Sudarsky, and Bill
Hubbard for help, advice, and useful discussions. 
This work was supported in part by NASA grants NAG5-10760, NAG5-10629, NAG5-7499, and NAG5-7073
and with the help of JPL contract \# 1241086 under the NASA Navigator program.
Tables of our potential curves are available from the first author upon request.

\clearpage

\figcaption{
Relative orientations of the hydrogen molecule and the electron cloud of the sodium atom in
the ground (3s) and excited (3p$_x$, 3p$_y$, 3p$_z$) states.  For the C$_{2v}$ symmetry (H$_2$ orientation
perpendicular to H$_2$-Na line) of the Na+H$_2$ complex,
all the excited electronic states have different energies
E(3p$_x$) $<$ E(3p$_y$) $<$ E(3p$_z$), but for the C$_{\infty}$ symmetry (all atoms colinear, not shown) the $B_1$ and $B_2$ states are
degenerate.  The $B$ states map onto the red wing, while the $A$ state maps onto the blue wing.
The corresponding figure for potassium is quite similar, but involves 4s and 4p states.  
\label{fig:1}}

\figcaption{
A schematic of the relationship between the interparticle distance 
$R$, the excitation energy $\Delta V$, and the spectrum.
Large shifts away from the atomic line center are associated with small 
distances where the chemical interaction is strongest.  The rainbow satellite 
feature appears at the radius (frequency) at which the excitation energy 
is maximum. Satellites are observed in the systems of concern in this paper
only on the blue wings.
\label{fig:2}}

\figcaption{
Interaction potentials (in eV, relative to zero at infinity) for all states of the Na+H$_2$ and K+H$_2$ systems
as a function both of the separation ($R$, in \AA) and of the angle of orientation of the H$_2$ molecule
with respect to the line between the alkali metal atom and the molecule.  The similar 
electronic structures of Na and K lead to similar shapes and behaviors for the 
curves.  The $A_1$ states (including the ground state) are repulsive for all separations, while the excited $^2B$ states can 
be attractive for a range of separations and angles.  Also included (long dash) are the corresponding curves 
for perturbations by helium. 
\label{fig:3}}

\figcaption{
Energy shifts ($\Delta V$, in eV) relative to the ground states for the Na+H$_2$  and K+H$_2$
complexes as a function of separation ($R$, in \AA).  The energy differences for the transitions to 
the excited $^2P_{3/2}$ ($^2B$) states are below those for the isolated D atomic lines (at 2.1 eV for Na and 1.6 eV for K)
and, hence, are associated with the red wings of the spectra.  The corresponding curves for the excited $^2P_{1/2}$ ($^2A$) states
are above those for the isolated D lines and, hence, are associated with the blue wings of the absorption spectra.
The extrema seen in the bottom panels associated with the blue wings mean that the blue wings
have satellites.  The corresponding curves for perturbations by helium are also shown (long dashed).
\label{fig:4}}

\figcaption{
Absorption cross sections (in cm$^2$) versus wavelength (in \mic) for the
Na-D and K I doublets at 5890 \AA\ and 7700 \AA, respectively, using the quasi-static
theory of wing line profiles.  The Lorentzian line cores are not included on this plot.  Cross sections
are shown for the different orientations of the H$_2$ molecule 
($\theta = 0^{\circ}, 20^{\circ}, 45^{\circ}, 70^{\circ}, 90^{\circ}$), as well as after
integrating over angle (solid).   The behavior is monotonic with angle, from steepest (small angle)
to shallowest (large angle).  Also depicted are the cross sections due to
perturbations by the spherical helium atom (with C$_{\infty}$ symmetry).
The asterisks indicate the positions of the rainbow satellites on the blue wings.
Due to resolution problems with the potential calculations (which are amplified
when using eq. (\ref{rainbow}) for which the second derivative $V^{\prime\prime}$ 
is needed), we have not attempted to obtain the precise shape of the satellite features.
A temperature of 1000 K and a total pressure of one atmosphere have been used.
The partial pressure of He is assumed to be $\sim$10\% of this total (Anders and Grevesse 1989). 
\label{fig:5}}

\figcaption{
Representative temperature and pressure dependences of the 
absorption cross sections for the Na/K + H$_2$ systems.  At one atmosphere pressure,
cross sections (in cm$^{2}$) at temperatures of 800 K (blue), 1000 K (red), and 1500 K (green) are depicted.  Another
curve at a pressure of 10 atmospheres and a temperature of 1000 K is shown (purple).  The left ends terminate at the 
central positions of the satellites.  Note that the positions of these rainbow features
are very weak functions of pressure and temperature and that the cutoff wavelength 
on the red wings (however defined) is an increasing function of temperature.    
\label{fig:6}}

\figcaption{
A comparison at a temperature of 1000 K and a pressure of one atmosphere of
the potassium absorption cross sections used in Burrows, Marley, and Sharp (BMS, 2000) (red) 
and Burrows et al. (2002) (green) with
those derived in this paper using the UFC theory (blue).  For the BMS curve, a $q$
parameter equal to 0.5 is used.  The dashed curve (blue) 
is the absorption cross section derived neglecting electron correlation (configuration
interaction, CI) effects.  The asterisk near 0.66 \mic marks the approximate position of the
K+H$_2$ satellite. 
\label{fig:7}}

\end{document}